\documentclass[a4paper,11pt]{amsart}
\begin{document}
\hyphenation{gra-vi-ta-tio-nal re-la-ti-vi-ty Gaus-sian
re-fe-ren-ce re-la-ti-ve gra-vi-ta-tion Schwarz-schild
ac-cor-dingly gra-vi-ta-tio-nal-ly re-la-ti-vi-stic pro-du-cing
de-ri-va-ti-ve ge-ne-ral ex-pli-citly des-cri-bed ma-the-ma-ti-cal
de-si-gnan-do-si coe-ren-za pro-blem gra-vi-ta-ting geo-de-sic
per-ga-mon cos-mo-lo-gi-cal gra-vity cor-res-pon-ding
de-fi-ni-tion phy-si-ka-li-schen ma-the-ma-ti-sches ge-ra-de
Sze-keres con-si-de-red tra-vel-ling}
\title[On the radial geodesics of Kerr's manifold]
{{\bf On the radial geodesics of Kerr's manifold}}

\author[Angelo Loinger]{Angelo Loinger}
\address{A.L. -- Dipartimento di Fisica, Universit\`a di Milano, Via
Celoria, 16 - 20133 Milano (Italy)}
\author[Tiziana Marsico]{Tiziana Marsico}
%\date{}
\address{T.M. -- Liceo Classico ``G. Berchet'', Via della Commenda, 26 - 20122 Milano (Italy)}
\email{angelo.loinger@mi.infn.it} \email{martiz64@libero.it}
%\thanks{}

\vskip0.50cm

\begin{abstract}
The so-called stationary-limit surface of Kerr's manifold
represents an insuperable barrier for the arriving material
elements and light-rays.
\end{abstract}

\maketitle

%%\begin{equation} \label{eq:sevenprime}
%%    \ddot{\Re} + \f\textbf{5}. External
%% {\kappa}{6}\Re \rho=0 , \tag{7'}
%% \end{equation}
%% ``mechanisms'' \textrm{d} \`a
%% \cite{1}
%% eqs.(\ref{eq:six})

\normalsize \vskip0.60cm \noindent \textbf{\emph{Summary.}} --
Introduction. -- \textbf{1}. Different forms of Kerr's interval
$\textrm{d}s^{2}$. -- \textbf{2}. A test-particle or a light-ray,
moving along a radial geodesic in the negative direction of the
radial coordinate, arrive at the ``stationary-limit'' surface with
a zero velocity and a positive, or zero, acceleration. --
\textbf{3}. Action of Hilbert's repulsive effect on the radial
geodesics for which $\vartheta=\pi/2$. -- \textbf{4}. Action of
Hilbert's repulsive effect on the radial geodesics for which
$\vartheta$ has a generic value.-- Appendix.


\normalsize

\vskip1.20cm \noindent \textbf{\emph{Introduction.}} -- At the end
of a paper of ours \cite{1} you find the intent to give in a
future work a reasonable explanation of those observational data,
reported by Remillard \emph{et al.} \cite{2} and by McClintock
\cite{2}, which suppositionally concern the event horizons of
Kerr's mass points \cite{3}, \cite{4}. Well, we do not give here
the promised explanation, we give instead a demonstration that
\emph{Kerr's horizons are incapable of swallowing anything}. The
reason of this fact is simple: as we shall see, the
stationary-limit surface of Kerr's manifold, which is external to
the event horizons, exerts a gravitational \emph{repulsion} on the
arriving material elements and light-rays, that is quite similar
to the Schwarzschildian repulsive action \cite{5}.

\vskip1.20cm \noindent \textbf{1.} -- As the most adequate choice
from the \emph{physical} standpoint, we shall choose the time
$t=x^{4}$ as evolution parameter \cite{6}, and we shall compute
velocity $\textrm{d}r/\textrm{d}t\equiv \dot{r}(t)$ and
acceleration $\textrm{d}^{2}r/\textrm{d}t^{2}\equiv \ddot{r}(t)$
of the test-particles and of the light-rays travelling along the
radial geodesics of Kerr's manifold.

\par It is however suitable to recall, first of all, some
important properties of the coordinate frames describing the above
manifold. With Boyer and Lindquist \cite{3}, we start from the
following form of Kerr's interval:

\begin{eqnarray} \label{eq:one}
\textrm{d}s^{2} & = & \textrm{d}x^{2} + \textrm{d}y^{2} +
\textrm{d}z^{2} - \textrm{d}t^{2} + \nonumber\\
& & {} %%
+ \frac{2mr^{3}}{r^{4}+a^{2}z^{2}} \times
\left[\frac{r(x\textrm{d}x+y\textrm{d}y)+a(x\textrm{d}y-y\textrm{d}x)}{r^{2}+a^{2}}
+ \frac{z\textrm{d}z}{r} + \textrm{d}t \right]^{2} \quad,
\end{eqnarray}

where $r$ is \emph{defined} by

\begin{equation} \label{eq:oneprime}
\left[(x^{2}+y^{2})  \, /  \, (r^{2}+a^{2})\right] + z^{2} / \,
r^{2} = 1 \quad; \tag{1'}
\end{equation}

here: $c=G=1$; $m$ and $a$ are two constants. The case $m\geq a$,
with $m$ and $a$ nonnegative parameters, is particularly
interesting. If $m=0$, the above $\textrm{d}s^{2}$ reduces to the
customary expression of Minkowskian interval. If $m\neq 0$ and
$a=0$, eqs. (\ref{eq:one}) - (\ref{eq:oneprime}) reduce to the
Eddington-Finkelstein form of the solution to Schwarzschild
problem of a gravitating point mass $m$ \cite{7}:

\begin{equation} \label{eq:two}
\textrm{d}s^{2} = \textrm{d}r^{2} + r^{2}
(\textrm{d}\vartheta^{2}+ \sin^{2}\vartheta \,
\textrm{d}\varphi^{2}) - \textrm{d}t^{2} + (2m/ \, r) \,
(\textrm{d}r + \textrm{d}t)^{2} \quad;
\end{equation}

remark that this interval was obtained from the standard (Hilbert,
Droste, Weyl) form of $\textrm{d}s^{2}$ with a coordinate
transformation whose derivative is singular at the value $2m$ of
standard $r$ ``in just the appropriate way for providing a
transformed metric that is regular there.'' \cite{8}.

\par A different form of Kerr's $\textrm{d}s^{2}$  is obtained by
putting, see \emph{e.g.} Boyer and Lindquist \cite{3}:

\begin{displaymath} \label{eq:three}
\left\{ \begin{array}{l}
x = \left(r^{2}+a^{2}\right)^{1/2} \, \sin\vartheta \, \cos \left[\varphi - \arctan(a/ \, r)\right] \quad, \\
y = \left(r^{2}+a^{2}\right)^{1/2} \, \sin\vartheta \, \sin \left[\varphi - \arctan(a/ \, r)\right] \quad, \\
z = r \cos\vartheta \quad; \tag{3}
\end{array} \right.
\end{displaymath}

the function $r$ of eqs. (\ref{eq:oneprime}) has been promoted to
the rank of a coordinate. If $\Sigma\equiv r^{2} + a^{2}
\cos^{2}\vartheta$, we get:

\setcounter{equation}{3}
\begin{eqnarray} \label{eq:four}
\textrm{d}s^{2} & = & \textrm{d}r^{2} + 2a \sin^{2}\vartheta \,
\textrm{d}r \, \textrm{d}\varphi + (r^{2}+a^{2}) \sin^{2}\vartheta
\, \textrm{d}\varphi^{2} + \Sigma(r,\vartheta) \,
\textrm{d}\vartheta^{2}
- \textrm{d}t^{2} + \nonumber\\
& & {} \quad \, \, + \left[2mr \, / \, \Sigma(r,\vartheta)\right]
\, (\textrm{d}r + a \sin^{2}\vartheta \, \textrm{d}\varphi +
\textrm{d}t)^{2} \quad.
\end{eqnarray}

If we wish a form of $\textrm{d}s^{2}$  such that for $a=0$ the
space coordinates reduce to the usual polar coordinates and the
interval reduces to the standard form of solution to Schwarzschild
problem, we must perform -- with Boyer and Lindquist \cite{3} --
the following substitutions, where $\Delta(r)\equiv r^{2}- 2mr +
a^{2}$:

\begin{displaymath} \label{eq:five}
\left\{ \begin{array}{l}
r \rightarrow r \quad; \quad \vartheta \rightarrow \vartheta \quad, \\
\textrm{d}\varphi \rightarrow  \textrm{d}\varphi - a \, \textrm{d}r / \Delta(r) \quad, \\
\textrm{d}t \rightarrow  \textrm{d}t + 2 \, m \, r \, \textrm{d}r
/ \Delta(r) \quad; \tag{5}
\end{array} \right.
\end{displaymath}
accordingly:

\setcounter{equation}{5}
\begin{eqnarray} \label{eq:six}
\textrm{d}s^{2} & = & \Sigma \, (\textrm{d}r^{2} / \, \Delta +
\textrm{d}\vartheta^{2}) + (r^{2}+a^{2}) \sin^{2}\vartheta \,
\textrm{d}\varphi^{2} - \textrm{d}t^{2} + \nonumber\\
& & {} + \left(2mr \, / \, \Sigma\right) \, (a \sin^{2}\vartheta
\, \textrm{d}\varphi + \textrm{d}t)^{2} \quad.
\end{eqnarray}

This $\textrm{d}s^{2}$ has two ``soft'' singularities at $r=m\pm
(m^{2}-a^{2})^{1/2}$ (``event horizons''). Two remarks: \emph{i})
the presence of mass $m$ in formulae (\ref{eq:five}) is
indispensable for the passage from an ``Eddington-Finkelstein''
coordinate frame (eqs. (\ref{eq:one}) - (\ref{eq:oneprime}) and
(\ref{eq:four})) to a ``standard'' one (eq. (\ref{eq:six}));
\emph{ii}) for $m=0$ and $a\neq 0$, we have, of course, a
Minkowskian $\textrm{d}s^{2}\equiv \textrm{d}s^{2}_{M}$:

\begin{eqnarray} \label{eq:seven}
\textrm{d}s^{2}_{M} & = &
\frac{r^{2}+a^{2}\cos^{2}\vartheta}{r^{2}+ a^{2}} \,
\textrm{d}r^{2} +  \, (r^{2}+a^{2}\cos^{2}\vartheta) \,
\textrm{d}\vartheta^{2}+ \nonumber\\
& & {} + (r^{2}+a^{2}) \sin^{2}\vartheta \, \textrm{d}\varphi^{2}
- \textrm{d}t^{2}   \quad;
\end{eqnarray}

it is not difficult to give direct proofs of the Minkowskian
character of (\ref{eq:seven}) (see Appendix). For $a\rightarrow
0$, $\textrm{d}s^{2}_{M}\rightarrow \textrm{d}r^{2} + r^{2}
(\textrm{d}\vartheta^{2}+ \sin^{2}\vartheta \,
\textrm{d}\vartheta^{2\mathcal{}}) - \textrm{d}t^{2}$. (Analogous
considerations hold for the form of $\textrm{d}s^{2}_{M}$ which
follows from eq. (\ref{eq:four}) when $m=0$.

\par The Minkowskian nature of Kerr's $\textrm{d}s^{2}$ when $m=0$
and $a\neq 0$ is not a trivial fact, because it proves that the
parameter $a$ does \emph{not} create a spacetime curvature
\emph{of its own}, but it is only a characteristic of some
\emph{reference frames}. Accordingly, it could be ``absorbed'' by
a suitable coordinate transformation.

\vskip1.20cm \noindent \textbf{2.} -- Let us consider the form
(\ref{eq:six}) of Kerr's interval. The generic radial geodesic of
a test-particle or of a light-ray is characterized by the
following formulae for the velocity $\textrm{d}r / \,
\textrm{d}t\equiv \dot{r}(t)$ and the acceleration
$\textrm{d}^{2}r / \textrm{d}t^{2}\equiv \ddot{r}(t)$, where the
constant $A$ is negative for the test-particles and zero for the
light-rays:

\begin{eqnarray} \label{eq:eight}
\dot{r}^{2} & = & \frac{\left( r^{2}+a^{2}\cos^{2}\vartheta - 2 \,
m \, r \right) \left( r^{2}+a^{2}-2 \, m \, r\right)}
{\left(r^{2}+
a^{2}\cos^{2}\vartheta\right)^{2}} -  \nonumber\\
& & {} - |A| \, \,  \frac{\left(r^{2}+a^{2}\cos^{2}\vartheta - 2
\, m \, r \right)^{2} \left(r^{2}+a^{2}-2mr\right)}{\left(r^{2}+
a^{2}\cos^{2}\vartheta\right)^{3}}  \quad;
\end{eqnarray}

the two roots of $r^{2}+a^{2}\cos^{2}\vartheta - 2mr=0$ are
$r_{1,2}:=m\pm (m^{2}-a^{2}\cos^{2}\vartheta)^{1/2}$. We have:

\begin{equation} \label{eq:nine}
\left[\dot{r}^{2}\right]_{r=r_{1,2}}=0 \qquad; \qquad
\left[\dot{r}^{2}\right]_{r=\infty}=1- |A| \quad.
\end{equation}

\begin{eqnarray} \label{eq:ten}
\ddot{r} & = & \frac{(r-m) \, [ \left(
r^{2}+a^{2}\cos^{2}\vartheta - 2 \, m \, r \right) + \left(
r^{2}+a^{2}-2 \, m \, r\right) ]} {\left(r^{2}+
a^{2}\cos^{2}\vartheta\right)^{2}} - \nonumber\\
& & {}
 \, \,  - \frac{2 \, r \, \left(r^{2}+a^{2}\cos^{2}\vartheta - 2 \,
m \, r \right) \left(r^{2}+a^{2}-2mr\right)}{\left(r^{2}+
a^{2}\cos^{2}\vartheta\right)^{3}} - \nonumber\\
& & {}
\, \, - |A| \, \, \left\{ \frac{2 \, (r-m) \,
(r^{2}+a^{2}\cos^{2}\vartheta - 2 \, m \, r )
(r^{2}+a^{2}-2mr)}{(r^{2}+
a^{2}\cos^{2}\vartheta)^{3}} \right.+ \nonumber\\
& & {} %%
\, \, + \frac{(r-m) \, (r^{2}+a^{2}\cos^{2}\vartheta - 2 \, m \, r
)^{2}}{(r^{2}+
a^{2}\cos^{2}\vartheta)^{3}} - \nonumber\\
& & {}
\, \, - \left. \frac{3 \, r \, (r^{2}+a^{2}\cos^{2}\vartheta - 2
\, m \, r )^{2} (r^{2}+a^{2}-2mr)}{(r^{2}+
a^{2}\cos^{2}\vartheta)^{4}}
 \right\} \quad.
\end{eqnarray}

For the acceleration $\ddot{r}(t)$ we have:

\begin{equation} \label{eq:eleven}
\left[\, \ddot{r} \, \right]_{r=r_{1,2}} = \pm
\frac{a^{2}\sin^{2}\vartheta \, \left( m^{2}- a^{2}
\cos^{2}\vartheta\right)^{1/2}}{\left\{2m \, [\, m  \pm
(m^{2}-a^{2}\cos^{2}\vartheta)^{1/2}]\right\}^{2}}\quad; \tag{11}
\end{equation}

\begin{equation} \label{eq:elevenprime}
\left[\, \ddot{r} \, \right]_{r=\infty}=0 \quad. \tag{11'}
\end{equation}

A test-particle or a light-ray, which arrives from the exterior
region, travelling in the negative direction of $r$, arrives at
the surface $r=r_{1}\equiv m +
(m^{2}-a^{2}\cos^{2}\vartheta)^{1/2}$; now:

\begin{equation} \label{eq:twelve}
\left[\, \ddot{r} \, \right]_{r=r_{1}} =
\frac{a^{2}\sin^{2}\vartheta \, \left( m^{2}- a^{2}
\cos^{2}\vartheta\right)^{1/2}}{\left\{2m \, [\, m  +
(m^{2}-a^{2}\cos^{2}\vartheta)^{1/2}]\right\}^{2}} \quad; \tag{12}
\end{equation}

this acceleration is positive (or zero for $\vartheta=0$ and
$\vartheta=\pi$); \emph{i.e.}, at $r=r_{1}$ we have a
gravitational \emph{repulsion}.

\par Thus, with a zero velocity and a positive (or zero)
acceleration no test-particle and no light-ray can cross the
barrier $r=r_{1}$.

\par Kerr's mass-point -- exactly as Schwarzschild's mass point
-- does not possess any marvellous property. By virtue, \emph{in
particular}, of the Hilbertian repulsive effect \cite{5}.

\vskip1.20cm \noindent \textbf{3.} -- The surface $r=r_{1}\equiv m
+ (m^{2}-a^{2}\cos^{2}\vartheta)^{1/2}$ is tangent to the surface
$r=r^{*}\equiv m + (m^{2}-a^{2})^{1/2}$ at $\vartheta=0$ and
$\vartheta=\pi$. For $\vartheta=\pi /2$, we have
$r_{1}[\vartheta=\pi /2]= 2m$; this case is particularly
interesting.

\par With a rather tedious computation, we find that
$\left(\dot{r}^{2}\right)_{b}$, the value of $\dot{r}^{2}$ which
corresponds to a value zero of the acceleration $\ddot{r}$
(attraction and repulsion \emph{balance} each other), is given by

\begin{equation} \label{eq:thirteen}
\left(\dot{r}^{2}\right)_{b} \, [\vartheta=\pi/2] = \frac{m \,
(r-2m)\,(r^{2}+a^{2}-2mr)^{2}}{r^{3}\,[\,3\,m\,r^{2}-(6m^{2}+a^{2})
\, r + 4ma^{2}]}\quad; \tag{13}
\end{equation}

for $a=0$, the right side of eq. (\ref{eq:thirteen}) yields
Hilbert's result \cite{5}: $(1/3) \, [(r- - 2m)/r]^{2}$.

\par Where $\dot{r}^{2}
<\left(\dot{r}^{2}\right)_{b}$ we have a gravitational
\emph{attraction}. Where $\dot{r}^{2}
>\left(\dot{r}^{2}\right)_{b}$ we have a gravitational
\emph{repulsion}.

\par For the light-rays we get from eq. (\ref{eq:eight}) -- with
$A=0$ -- that

\begin{equation} \label{eq:fourteen}
\dot{r}^{2} \, [\vartheta=\pi/2] =
\frac{(r-2m)\,(r^{2}+a^{2}-2mr)}{r^{3}} \, \geq \,
\left(\dot{r}^{2}\right)_{b} \, [\vartheta=\pi/2] \quad; \tag{14}
\end{equation}

\emph{i.e.}, the light-rays are repulsed for \emph{all} values of
$r\geq 2m$. An analogous result was found by Hilbert \cite{5} for
Schwarzschild manifold. (The region $r<2m$ is quite unphysical).

\vskip1.20cm \noindent \textbf{4.} -- For a generic value of
$\vartheta$ we have results quite similar to those of sect. 3. We
write here the expression of $\left(\dot{r}^{2}\right)_{b}$:

\setcounter{equation}{14}
\begin{eqnarray} \label{eq:fifteen}
\left(\dot{r}^{2}\right)_{b} & = & \left[ \, m \,
(r^{2}+a^{2}-2mr)^{2} \, (r^{2}+a^{2}\cos^{2}\vartheta-2mr)\, (r^{2}-a^{2}\cos^{2}\vartheta) \right] \times \nonumber\\
& & {}
\times \left\{ (r^{2}+a^{2}\cos^{2}\vartheta)^{2} \, [a^{4} r
\cos^{2}\vartheta \, (\cos^{2}\vartheta-1) \, - \right. \nonumber\\
& & {}
- a^{4}m\cos^{2}\vartheta \, (\cos^{2}\vartheta+2)
+(a^{2}r^{3}-4a^{2}r^{2}m) \, (\cos^{2}\vartheta-1) + \nonumber\\
& & {}
\left. + \, 6a^{2}m^{2}r \cos^{2}\vartheta + 3r^{4}m -
6r^{3}m^{2}] \right\}^{-1} \quad;
\end{eqnarray}

for $a=0$, we obtain Hilbert's result \cite{5}, \emph{i.e.} $(1/3)
\, [(r-2m)/r]^{2}$. Where $\dot{r}^{2}
<\left(\dot{r}^{2}\right)_{b}$ we have a gravitational
\emph{attraction}. Where $\dot{r}^{2}
>\left(\dot{r}^{2}\right)_{b}$ we have a gravitational
\emph{repulsion}. The light-rays are repulsed for \emph{all}
values of $r\geq m + (m^{2}+a^{2}\cos^{2}\vartheta)^{1/2}$.

\par There is an analogous result for Schwarzschild's manifold for
$r\geq 2m$ \cite{5}.

\par A last remark. The repulsive effect makes evident that for
Schwarzschild's and Kerr's manifolds the question of
\emph{geodesic completeness}, though interesting from a geometric
standpoint, is destitute of a physical importance.

\vskip1.20cm
\par \emph{Acknowledgment}. We are grateful to Dr.
Giovanni Rastelli, who has obtained with a computer program
formulae (\ref{eq:thirteen}) and (\ref{eq:fifteen}), and the
analytical proof of the Appendix.

\newpage
% \vskip2.00cm
\begin{center}
\noindent \small \emph{\textbf{APPENDIX}}
\end{center}
\normalsize \noindent \vskip0.80cm

\par The metric tensor of eq. (\ref{eq:seven}) is regular, time
independent, and satisfies $R_{jk}=0$. Consequently, by virtue of
Serini's theorem, it satisfies also $R_{jklm}=0$, \emph{i.e.} the
$\textrm{d}s^{2}_{M}$ of eq. (\ref{eq:seven}) has a Minkowskian
character.

\par An analytical proof: we start from $\textrm{d}s^{2}_{M}= \textrm{d}x^{2} + \textrm{d}y^{2} + \textrm{d}z^{2} -
\textrm{d}t^{2}$; the transformations

\begin{displaymath} \label{eq:A1}
\left\{ \begin{array}{l}
x = r' \sin\vartheta'  \, \cos\varphi' \quad, \\
y = r' \sin\vartheta'  \, \sin\varphi' \quad, \\
z = r' \cos\vartheta' \tag{A1}
\end{array} \right.
\end{displaymath}

give, as it is well known:

\begin{equation} \label{eq:A2}
\textrm{d}s^{2}_{M} = \textrm{d}r'^{2} + r'^{2} \,
(\textrm{d}\vartheta'^{2}+ \sin^{2}\vartheta' \,
\textrm{d}\varphi'^{2}) - \textrm{d}t^{2}  \quad; \tag{A2}
\end{equation}

the transformations:

\begin{displaymath} \label{eq:A3}
\left\{ \begin{array}{l}
x = (r^{2}+a^{2})^{1/2} \, \sin\vartheta  \, \cos\varphi \quad, \\
y = (r^{2}+a^{2})^{1/2} \, \sin\vartheta  \, \sin\varphi \quad, \\
z = r \cos\vartheta \tag{A3}
\end{array} \right.
\end{displaymath}

give eq. (\ref{eq:seven}), \emph{i.e.}:

\begin{equation} \label{eq:A4}
\textrm{d}s^{2}_{M} = \frac{r^{2}+a^{2}\,
\cos^{2}\vartheta}{r^{2}+a^{2}} \, \textrm{d}r^{2} +
(r^{2}+a^{2}\, \cos^{2}\vartheta)  \, \textrm{d}\vartheta^{2} +
(r^{2}+a^{2}) \, \sin^{2}\vartheta \, \textrm{d}\varphi^{2} -
\textrm{d}t^{2} \quad.    \tag{A4}
\end{equation}

\qquad \qquad \qquad \qquad \qquad \qquad \qquad \qquad \qquad
\qquad \qquad \qquad \qquad \qquad \emph{Q.e.d.}

The relation between $(r', \vartheta', \varphi')$ and $(r,
\vartheta, \varphi)$ is:

\begin{displaymath} \label{eq:A5}
\left\{ \begin{array}{l}
r'^{2} = r^{2}+a^{2} \, \sin^{2}\vartheta  \quad, \\
\cos^{2}\vartheta' = \left(r^{2} \, cos^{2}\vartheta\right)\, / \,  \left(r^{2} + a^{2} \, sin^{2}\vartheta\right) \quad, \\
\tan\varphi' = \tan\varphi \tag{A5}
\end{array} \right.
\end{displaymath}

(The Minkowskian character of Kerr's interval for $m=0$ and $a\neq
0$ is quite evident from eqs. (\ref{eq:one}) --
(\ref{eq:oneprime}).)


\vskip1.80cm \small
\begin{thebibliography}{9}

\bibitem{1}
A. Loinger and T. Marsico: \emph{a}) \emph{arXiv:0706.3891 v3}
$[$physics.gen-ph$]$ 16 Jul 2007.

\bibitem{2}
R. Remillard \emph{et al.}, \emph{arXiv:astro-ph/0208402 v1}
(August 21tst, 2002); J. E. McClintock, \emph{Astrophys. J.},
652518 (2006-06-04).

\bibitem{3}
See, in particular: R.P. Kerr, \emph{Phys. Rev. Letters},
\textbf{11} (1963) 237; R.H. Boyer and R.W Lindquist, \emph{J.
Math. Phys.}, \textbf{8} (1967) 265; C. M\o ller, \emph{The Theory
of Relativity}, Second Edition (Clarendon Press, Oxford) 1972, p.
482 \emph{sqq.}; S. Chandrasekhar, \emph{The Observatory},
\textbf{92} (1972) 160.

\bibitem{4}
A. Loinger, \emph{arXiv:gr-qc/9911077} (January 5th, 2000); it is
proved here that simple charges of the radial coordinate leave no
room for the attribution of extraordinary properties to the field
of a Kerr's corpuscle.

\bibitem{5}
D. Hilbert, \emph{Mathem. Annalen}, \textbf{92} (1924) 1; also in
\emph{Gesammelte Abhandlungen}, Dritter Band (J. Springer, Berlin)
1935, p.258.

\bibitem{6}
Cf., \emph{e.g.}, V. Fock, \emph{The Theory of Space, Time and
Gravitation}, Second Revised Edition (Pergamon Press, Oxford,
\emph{etc}) 1964, pp. 134 and 135.

\bibitem{7}
A.S. Eddington, \emph{Nature}, \textbf{113} (1924) 192; D.
Finkelstein, \emph{Phys. Rev.}, \textbf{110} (1958) 965.

\bibitem{8}
S. Antoci and D.-E. Liebscher, \emph{Astr. Nachrichten},
\textbf{322} (2001) 137.

\end{thebibliography}
\end{document}